\documentclass[sn-aps]{sn-jnl}

\usepackage{graphicx}%
\usepackage{multirow}%
\usepackage{amsmath,amssymb,amsfonts}%
\usepackage{amsthm}%
\usepackage{mathrsfs}%
\usepackage[title]{appendix}%
\usepackage{xcolor}%
\usepackage{textcomp}%
\usepackage{manyfoot}%
\usepackage{booktabs}%
\usepackage{algorithm}%
\usepackage{algorithmicx}%
\usepackage{algpseudocode}%
\usepackage{listings}%

\theoremstyle{thmstyleone}%
\theoremstyle{thmstyletwo}%

\theoremstyle{thmstylethree}%

\begin{document}
\title[Fluctuations in the email size modeled by a log-normal-like distribution]{Fluctuations in the email size modeled by a log-normal-like distribution}


\author*[1]{\fnm{Yoshitsugu} \sur{Matsubara}}\email{matubara@cc.saga-u.ac.jp}



\affil*[1]{\orgdiv{Computer and Network Center}, \orgname{Saga University}, \orgaddress{\street{1 Honjo-machi}, \city{Saga-shi}, \postcode{840-8502}, \state{Saga}, \country{Japan}}}




\abstract{
A previously established frequency distribution model combining a log-normal distribution with a logarithmic equation describes fluctuations in the email size during send requests.
Although the frequency distribution fit was considered satisfactory, the underlying mechanism driving this distribution remains inadequately explained.
To address this gap, this study introduced a novel email-send model that characterizes the sending process as an exponential function modulated by noise from a normal distribution.
This model is consistent with both the observed frequency distribution and the previously proposed frequency distribution model.
}

\keywords{Email size, Power-law fluctuations, Log-normal-like distribution}

\maketitle
\section{Introduction} \label{sec1}

The complex structures and dynamic behaviors of internet systems have attracted considerable attention.
Phenomena such as scale-free structures~\cite{Faloutsos:1999} and power-law distributions in packet flows~\cite{Paxson:1995,Csabai:1994,Takayasu:1996,Tadaki:2007,Eckmann:2004,Barabasi:2005,K.-I.:2008,Malmgren:2008,Anteneodo:2010,Karsai:2012} during inter-event times have been studied extensively.
Studies have investigated power-law correlations in data flow and email send requests~\cite{Matsubara:2013} during interevent times.

A previous study identified power-law fluctuations in the email size during send requests~\cite{Matsubara:2017} and proposed a novel model based on a log-normal-like distribution.
This distribution effectively captured the power-law characteristics of large email sizes and has been reported in numerous studies ~\cite{ASI:ASI21331}.
However, the mechanisms underlying log-normal-like distributions remain inadequately explained.
To address this gap, the current study proposed a novel email-size generation model to explain these mechanisms.
Simulation results from the proposed model demonstrate exhibit strong agreement with observed frequency distributions and the previously proposed frequency distribution model.
Furthermore, linguistic principles have been applied to explain the underlying mechanism of the model.

The paper is structured as follows:
Section~\ref{sec2} examines the log-normal-like distribution and its connections to relevant linguistic studies.
Section~\ref{propose} presents the proposed email-size generation model and evaluates its fit to both the observed frequency distribution and the frequency distribution model


\section{Email size fluctuations}\label{sec2}

This section summarizes the analysis of email size fluctuations presented in a previous study~\cite{Matsubara:2017}, which examined the frequency distribution of email sizes across various periods and user groups from May 1, 2015, to July 31, 2015.
Two inflection points were identified in the size-frequency distribution, approximately $15 \,\mathrm{kB}$ and $40 \,\mathrm{kB}$.
The analysis focused on emails with attachments, as plain text emails typically remained below a few tens of kilobytes.
The size-frequency distribution was categorized into two subdistributions based on content type, namely ``No attachment'' and ``Attachment,'' where ``No attachment'' refers to plain text and HTML emails.
Content types were defined according to the multipurpose internet mail extensions (MIME) protocol~\cite{rfc2045,rfc2046,rfc2047,rfc2049,rfc4289,rfc6838}.

Because email bodies consist of written sentences, linguistic principles were incorporated into the analysis.
Most users in the organization from which the data were collected were Japanese, and emails were predominantly written in Japanese and English.
Linguistic studies analyzing sentence length have identified several distribution types, including log-normal~\cite{Arai:2001,Furuhashi:2012}, gamma~\cite{Sasaki:1976}, and hyper-Pascal distributions~\cite{Ishida:2007}.

Based on this analysis, we proposed a novel model to explain the email size frequency distribution within each subdistribution.
The proposed model was evaluated for its goodness of fit with the observed data.

The model is defined by a probability density function $p(s)$ for each subdistribution of email size frequencies (where $s$ is measured in units of 100 bytes) as follows:
\begin{eqnarray}
p(s) & = & \frac{1}{a}\frac{1}{s\ln{s}} \exp{ \left\{ \frac{- (\ln{\ln{s}} - \mu)^{2}}{2 \sigma^{2}} \right\} } \label{model_eq_log-normal-like},
\end{eqnarray}
where $a$ is the normalized constant ($a > 0$), $-\infty < \mu < \infty$, and $\sigma > 0$\footnote{.and the function is continuous over the range ($1,\infty$) for $s$. Eq.~\ref{model_eq_log-normal-like} integrates the log-normal distribution $g(x) = \frac{1}{a}\frac{1}{x}\exp(-\frac{(\ln{x} - \mu)^{2}}{2\sigma^{2}})$ with $x = \ln{s}$.}.

The expected value of $p(s)$ remained undefined.
The logarithms in eq.~\ref{model_eq_log-normal-like} for $s \gg 1$ can be approximated as follows:
\begin{eqnarray}
\ln{p(s)} & \sim & - \ln{s} - \frac{(\ln{\ln{s}})^{2}}{2 \sigma^{2}} \nonumber \\
& \sim & - \ln{s}, \nonumber
\end{eqnarray}
where $g(x)$ does not exhibit power-law behavior for $x \gg 1$.
Consequently, $x = \ln{s}$ influences the power-law properties of $p(s)$ when $s \gg 1$.

The least-squares method was used to evaluate the goodness-of-fit of the proposed model to the observed data in accordance with a previous study~\cite{Matsubara:2017}.
The method is defined as follows:
\begin{equation}
D = \sum_{s} (\ln{y_{O}(s)} - \ln{y(s)})^2,
\end{equation}
where $y_{O}(s)$ represents the relative frequency of each bin size in the observed data and $y(s)$ corresponds to the relative frequency derived from the proposed model.
We used $p(s)$ as the relative frequency by adjusting the normalization constant a and the size range $s$ for each bin.
$D$ approaches zero when $y_{O}(s) \simeq y(s)$.
The parameter values $\mu$ and $\sigma$ of $p(s)$ that minimized $D$ were selected for the model.

\begin{figure}[t]
\centering
\resizebox{1.00\textwidth}{!}{\includegraphics{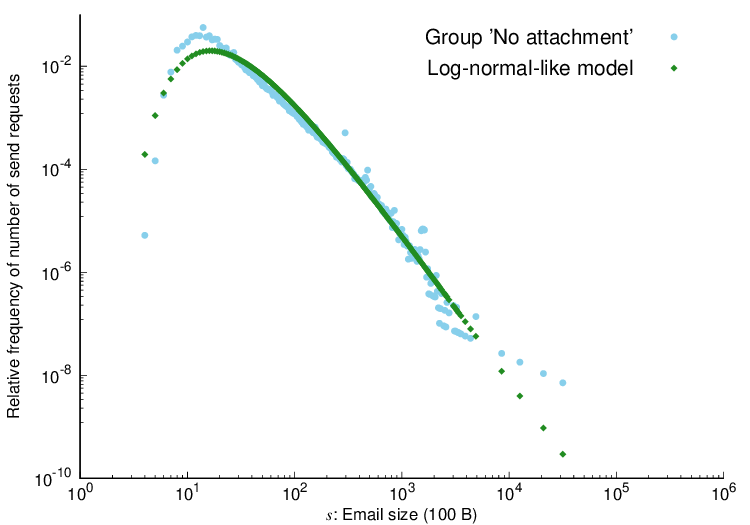}}
\caption{
Relative frequency distribution of the ``No attachment'' email group along with the corresponding log-normal-like fitted distribution (eq.~\ref{model_eq_log-normal-like}).
Both distributions are logarithmically binned~\cite{ASI:ASI21426}, and both axes are on a logarithmic scale.
The blue points represent the observed relative frequency values, whereas the green points indicate those calculated using $p(s)$.
The horizontal axis represents the email size $s$ (in units of 100 bytes), with the bin size $\Delta s$ defined as $(10^{0.01} - 1) s$.
All bin intervals have an equal logarithmic scale.
The vertical axis depicts the relative frequency of send requests
}
\label{res_size_frequency_per100B_noattachment_log-normal-like_log10}
\end{figure}

The fitted distribution obtained using the proposed model (eq.~\ref{model_eq_log-normal-like}) for the ``No attachment'' case is depicted in fig.~\ref{res_size_frequency_per100B_noattachment_log-normal-like_log10}.
The parameter values for the fitted curves were $\mu = 1.2599$ and $\sigma = 0.2461$.
The model fits the observed data well (fig.~\ref{res_size_frequency_per100B_noattachment_log-normal-like_log10}).

\section{Email size generation model} \label{propose}

Let $s_{t}$ denote the email size at time $t$.
The size is described by the following equation: 
\begin{equation}
s_{t} = b s_{t-1}^{c} \mathrm{e}^{\mathrm{e}^{\epsilon_{t}}}, \label{model_eq_email-size-generation}
\end{equation}
where $b$ and $c$ are coefficients and $\epsilon_{t}$ represents noise following a normal distribution $\mathcal{N}(\mu,\sigma^{2})$.

Equation ~\ref{model_eq_email-size-generation} can be interpreted as follows:
\begin{enumerate}

\item The email size is influenced by the length of the email body, which is proportional to the sentence length.
A direct relationship exists between the text length and email size.

\item The body content is written in natural languages such as English or Japanese.

\item $\epsilon_{t}$ corresponds to the length of a single word.

\item $\exp\{\epsilon_{t}\}$ corresponds to the length of a single sentence.
Each sentence includes at least one word, with word choice depending on the sentence’s meaning and structure.
Additive processes are unlikely because words are not selected randomly.
Therefore, $\exp\{\epsilon_{t}\}$ represents a multiplicative process~\cite{MOHR2022351:2022}.

\item $\exp\{\exp\{\epsilon_{t}\}\}$ corresponds to the length of a compound sentence.
A compound sentence consists of at least one single sentence.
Sentence selection also depends on meaning and structure, precluding independent random selection.
Therefore, the exponential form $\exp\{\exp\{\epsilon_{t}\}\}$ is used to model this multiplicative process.

\item $s_{t-1}^{c}$ represents the effect of prior emails.
Some emails contained quoted content from previous emails, referred to as the quotation effect.
If $c \simeq 0$, then $s_{t}$ is described as
\begin{equation}
s_{t} \simeq b \mathrm{e}^{\mathrm{e}^{\epsilon_{t}}} \label{model_eq_email-size-generation-simplest}
\end{equation}
as a simple case.
When $c \simeq 0$, most emails are independent of prior emails.
Although some emails contain quoted content, its inclusion has a negligible effect on the size-frequency distribution.

\end{enumerate}


\section{Discussion} \label{Discussion}

\begin{figure}[t]
\centering
\resizebox{1.00\textwidth}{!}{\includegraphics{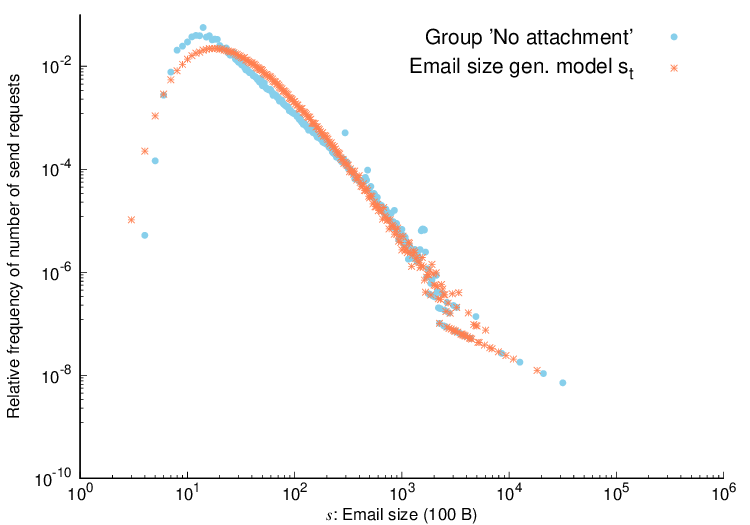}}
\caption{
Size-frequency distribution of ``No attachment'' emails in observed data and those generated by the email size generation model $s_{t}$.
The frequency distribution is logarithmically binned~\cite{ASI:ASI21426}, with both axes on a logarithmic scale.
The blue points represent the observed relative frequency values, whereas the orange points correspond to those calculated using $s_{t}$.
The horizontal axis indicates the email size $s$ (in units of 100 bytes), ranging from $0 \, \mathrm{MB}$ to $10 \, \mathrm{MB}$, with the bin size $\Delta s$ defined as $(10^{0.01} - 1) s$.
All bin intervals maintain a consistent size on the logarithmic scale.
The vertical axis denotes the relative frequency of emails.
Model parameters for $s_{t}$ are $b = 105$, $c = 0$, $\mu = 1.259$, and $\sigma = 0.235$.
The degree of fitting is $D = 16.055$
}
\label{res_size_frequency_per100B_noattachment_simulation_log10_1}
\end{figure}

\begin{figure}[t]
\centering
\resizebox{1.00\textwidth}{!}{\includegraphics{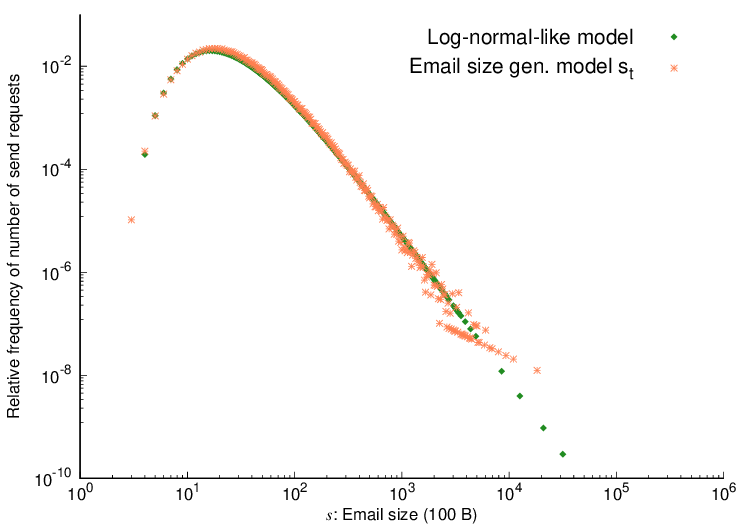}}
\caption{
Size-frequency distribution of ``No attachment'' emails in observed data and those generated by the email size generation model $s_{t}$.
The frequency distribution is logarithmically binned~\cite{ASI:ASI21426}, and both axes are on a logarithmic scale.
The green points represent those calculated using $p(s)$, whereas the orange points correspond to those calculated using $s_{t}$.
The horizontal axis depicts the email size $s$ (in units of 100 bytes), ranging from $0 \, \mathrm{MB}$ to $10 \, \mathrm{MB}$, with the bin size $\Delta s$ defined as $(10^{0.01} - 1) s$.
All bin intervals maintain a consistent size on the logarithmic scale.
The vertical axis denotes the relative frequency of emails.
Model parameters for $s_{t}$ are $b = 105$, $c = 0$, $\mu = 1.259$, and $\sigma = 0.235$.
The degree of fitting is $D = 4.294$
}
\label{res_size_frequency_per100B_noattachment_simulation_log10_2}
\end{figure}

The email size frequency distribution generated using st was simulated, and the results are presented in fig.~\ref{res_size_frequency_per100B_noattachment_simulation_log10_1} and fig.~\ref{res_size_frequency_per100B_noattachment_simulation_log10_2}.
This study generated 191,993 emails based on $s_{t}$, matching the number of ``No attachment'' emails analyzed in the previous study~\cite{Matsubara:2017}.
The parameter value of $c$ for $s_{t}$ was $c = 0$, which is the simplest case in eq.~\ref{model_eq_email-size-generation-simplest}.
The other parameter values for $s_{t}$ were $b = 105$, $\mu = 1.259$, and $\sigma = 0.235$.
The least-squares method denoted by $D$ was used to assess the goodness-of-fit.
The degree of fitting was each $D = 16.055$ and $D = 4.294$.
The frequency distribution of emails generated by $s_{t}$ revealed a strong fit to both the observed data and the email size frequency model $p(s)$.

%
%
%
%

%
%

\begin{figure}[t]
\centering
\resizebox{1.00\textwidth}{!}{\includegraphics{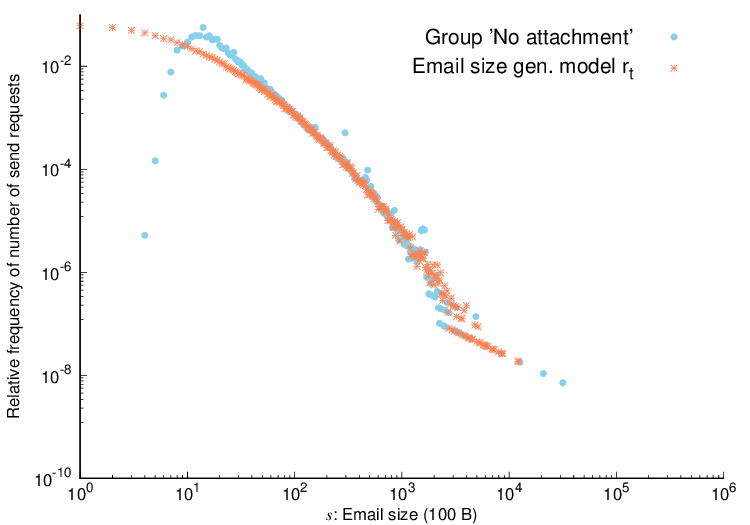}}
\caption{
Size-frequency distribution of ``No attachment'' emails in observed data and those generated by the email size generation model $r_{t}$.
This frequency distribution is logarithmically binned~\cite{ASI:ASI21426}, and both axes are on a logarithmic scale.
The blue points represent observed relative frequency values, whereas the orange points depict those calculated using $r_{t}$.
The horizontal axis displays the email size s (in units of 100 bytes), ranging from $0 \, \mathrm{MB}$ to $10 \, \mathrm{MB}$, with bin size $\Delta s$ defined as $(10^{0.01} - 1) s$.
All bin intervals are uniformly spaced on the logarithmic axis.
The vertical axis represents the relative frequency of emails.
Parameters for $r_{t}$ are $b = 87.9921$, $\mu = 2.7616$, and $\sigma = 1.5623$.
The degree of fitting is $D = 33.271$
}
\label{res_size_frequency_per100B_noattachment_simulation_ln_log10_1}
\end{figure}

\begin{figure}[t]
\centering
\resizebox{1.00\textwidth}{!}{\includegraphics{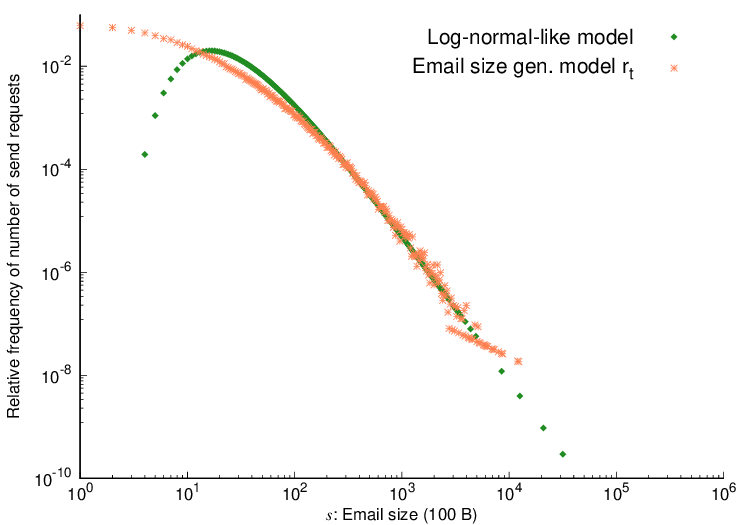}}
\caption{
Size-frequency distribution generated by the size-frequency distribution model $p(s)$ and the email size generation model $r_{t}$.
This frequency distribution is logarithmically binned~\cite{ASI:ASI21426}, and both axes are on a logarithmic scale.
The green points represent the relative frequency values calculated using $p(s)$, whereas the orange points indicate those calculated using $r_{t}$.
The horizontal axis represents email size $s$ (in units of 100 bytes), ranging from $0 \, \mathrm{MB}$ to $10 \, \mathrm{MB}$, with bin size $\Delta s$ defined as $(10^{0.01} - 1) s$.
All bin intervals have a uniform size on the logarithmic scale.
The vertical axis reveals the relative frequency of emails.
Parameters for the model $r_{t}$ are $b = 87.9921$, $\mu = 2.7616$, and $\sigma = 1.5623$.
The degree of fitting is  $D = 16.669$
}
\label{res_size_frequency_per100B_noattachment_simulation_ln_log10_2}
\end{figure}

Additionally, the frequency distribution of emails generated using $r_{t}$ was simulated for comparison.
The equation is described as follows:
\begin{equation}
r_{t} = b \mathrm{e}^{\epsilon_{t}}. \label{model_eq_email-size-generation-ln}
\end{equation}
The fitting results are displayed in fig.~\ref{res_size_frequency_per100B_noattachment_simulation_ln_log10_1} and fig.~\ref{res_size_frequency_per100B_noattachment_simulation_ln_log10_2}.
The parameter values for $r_{t}$ were $b = 87$, $\mu = 2.774$, and $\sigma = 1.562$.
The degree of fitting was each $D = 33.271$ and $D = 16.669$.
The fit obtained with $r_{t}$ was inferior to that of $s_{t}$.
Therefore, the form $\exp\{\exp\{\epsilon_{t}\}\}$ represents the log-normal-like distribution.

%
%
%
%

\begin{figure}[t]
\centering
\resizebox{1.00\textwidth}{!}{\includegraphics{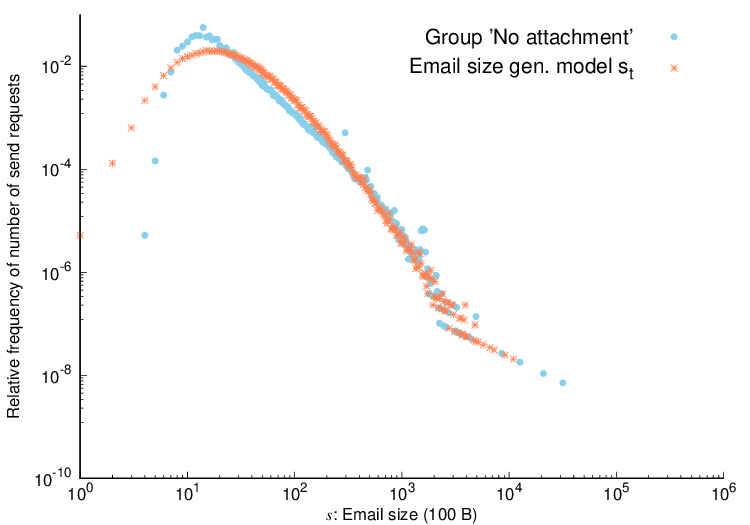}}
\caption{
Size-frequency distribution of ``No attachment'' emails in observed data and those generated by the email size generation model $s_{t}$ with  $c = 0.5$.
The frequency distribution is logarithmically binned~\cite{ASI:ASI21426}, with both axes on a logarithmic scale.
The blue points represent the observed relative frequency values, whereas the orange points correspond to those calculated using $s_{t}$.
The horizontal axis depicts the email size $s$ (in units of 100 bytes), ranging from $0 \, \mathrm{MB}$ to $10 \, \mathrm{MB}$, with the bin size $\Delta s$ defined as $(10^{0.01} - 1) s$.
All bin intervals maintain a consistent size on the logarithmic scale.
The vertical axis denotes the relative frequency of emails.
Parameters for the model $s_{t}$ are $b = 1.8$, $\mu = 1.259$, and $\sigma = 0.215$.
The degree of fitting is $D = 21.663$
}
\label{res_size_frequency_per100B_noattachment_simulation_log10_3}
\end{figure}

\begin{figure}[t]
\centering
\resizebox{1.00\textwidth}{!}{\includegraphics{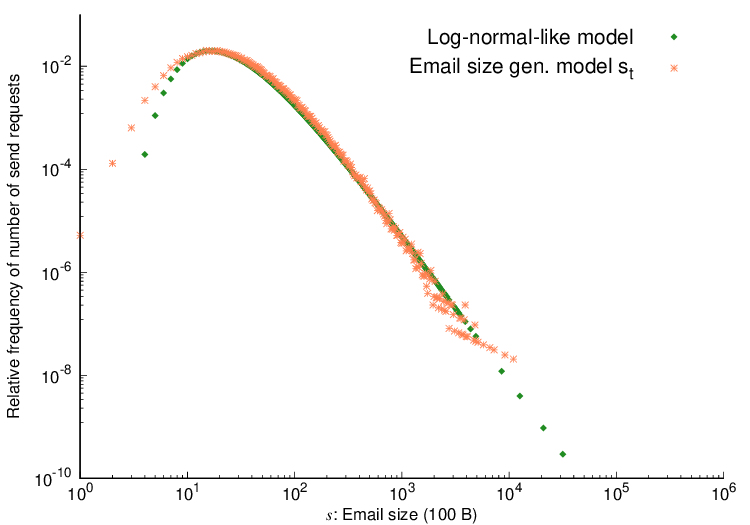}}
\caption{
Size-frequency distribution of ``No attachment'' emails in observed data and those generated by the email size generation model $s_{t}$ with $c = 0.5$.
The frequency distribution is logarithmically binned~\cite{ASI:ASI21426}, with both axes on a logarithmic scale.
The green points represent those calculated using $p(s)$, whereas the orange points correspond to those calculated using $s_{t}$.
The horizontal axis depicts email size $s$ (in units of 100 bytes) ranging from $0 \, \mathrm{MB}$ to $10 \, \mathrm{MB}$, with the bin size $\Delta s$ defined as $(10^{0.01} - 1) s$.
All bin intervals maintain a consistent size on the logarithmic scale.
The vertical axis denotes the relative frequency of emails.
Parameters for the model $s_{t}$ are $b = 1.8$, $\mu = 1.259$, and $\sigma = 0.215$.
The degree of fitting is $D = 5.362$
}
\label{res_size_frequency_per100B_noattachment_simulation_log10_4}
\end{figure}

Next, the fitting for $c = 0.5$ was simulated as an example of $c \neq 0$.
The fitting results are depicted in fig.~\ref{res_size_frequency_per100B_noattachment_simulation_log10_3} and fig.~\ref{res_size_frequency_per100B_noattachment_simulation_log10_4}.
The other parameter values for $s_{t}$ were $b = 1.8$, $\mu = 1.259$, and $\sigma = 0.215$.
The degree of fitting was each $D = 21.663$ and $D = 5.362$.
The fitting was better than that of $r_{t}$.
However, for $s_{t} < 1 \, \mathrm{KB}$, the fitting did not improve  compared with that of $s_{t} \, (c = 0)$.
These results suggests that most emails are independent of each other.

%
%
%
%


\section{Conclusion} \label{conclusion}

A previous study analyzed the frequency distributions of email sizes in sent requests~\cite{Matsubara:2017} and proposed a log-normal-like model that combined the log-normal distribution $g(x)$ and $x = \ln{s}$, where $s$ denotes the email size.
The proposed model approximates the power-law correlations when $s \gg 1$.
Although prior studies focused on explaining email size fluctuations, they did not sufficiently elucidate the mechanism underlying the generation of these frequency distributions.

To address this gap, this study proposed a novel email size generation model, $s_{t}$, to explain the mechanism of the size–frequency distribution $p(s)$.
The model equation, $s_{t} = b s_{t-1}^{c} \exp\{\exp\{\epsilon_{t}\}\}$, incorporates normal distribution noise $\epsilon_{t}$.
This model $s_{t}$ is explained based on linguistic principles.
The frequency distribution generated using $s_{t}$ demonstrated a strong fit to both the observed data and the email size frequency model $p(s)$.
Simulations with $s_{t}$ reveal that the log-normal-like distribution, derived from $\exp\{\exp\{\epsilon_{t}\}\}$, and the observed email data from the previous study indicate that most emails are independent of each other.


The normal distribution $\mathcal{N}(\mu, \sigma^{2})$ is used as noise $\epsilon_{t}$ in eq.~\ref{model_eq_email-size-generation}.
Here, $\epsilon_{t}$ corresponds to the length of a single word in a sentence.
The length of one word should be at least one character (one byte).
However, the normal distribution theoretically allows negative values.
This study explained the email size frequency distribution.
A simpler and more effective explanatory model than the current $s_{t}$ model should be developed using a different distribution in future research. 

\backmatter
\bmhead{Compliance with ethical standards}

The author has no competing interests to declare that are relevant to the content of this article.





\bmhead{Data Availability Statement}

The data are not publicly available to avoid compromising the privacy of the information. 

\bmhead{Acknowledgments}

The author thanks Makoto Otani, Computer and Network Center, Saga University, for assisting with data collection and analysis. 

\bibliography{matsubara,rfc}

\end{document}